\documentclass[conference,letterpaper]{IEEEtran}
\usepackage[utf8]{inputenc}
\usepackage{amsmath}
\usepackage{amssymb}
\usepackage{amsmath,amssymb,amsthm}
\usepackage{enumerate}
\usepackage{stfloats}
\usepackage{comment}
\usepackage{bbm}
\usepackage{subcaption}
\usepackage{siunitx}
\usepackage{mathtools}
\usepackage{accents,latexsym,cancel}
\usepackage{cite}

\usepackage[dvipsnames]{xcolor}
\definecolor{myPink}{RGB}{255,105,183}

\usepackage[T1]{fontenc}
\usepackage{graphics} 
\usepackage{epsfig} 
\usepackage[mathscr]{euscript}
\usepackage{algorithm}
\usepackage[noend]{algpseudocode}
\usepackage{bbm}
\makeatletter
\def\BState{\State\hskip-\ALG@thistlm}
\makeatother

\usepackage{tikz}
\usetikzlibrary{arrows,shapes,chains,matrix,positioning,scopes,patterns,calc,
decorations.markings,
decorations.pathmorphing,
}

\usepackage{pgfplots}
\pgfplotsset{compat=1.3}
\usepgflibrary{shapes}

\renewcommand{\epsilon}{\varepsilon}

\newcommand{\expt}{\mathbb{E}}

\newcommand{\RNum}[1]{\uppercase\expandafter{\romannumeral #1\relax}}

\newcommand{\av}{\ensuremath{\mathbf{a}}}

\newcommand{\cv}{\ensuremath{\mathbf{c}}}

\newcommand{\mv}{\ensuremath{\mathbf{m}}}

\newcommand{\wv}{\ensuremath{\mathbf{w}}}
\newcommand{\xv}{\ensuremath{\mathbf{x}}}
\newcommand{\yv}{\ensuremath{\mathbf{y}}}

\def\Pr{\mathrm{Pr}}

\DeclareMathAlphabet{\mcl}{OMS}{cmsy}{m}{n}

\newlength\tikzwidth
\newlength\tikzheight

\definecolor{mycolor1}{rgb}{0.63529,0.07843,0.18431}%
\definecolor{mycolor2}{rgb}{0.00000,0.44706,0.74118}%
\definecolor{mycolor3}{rgb}{0.00000,0.49804,0.00000}%
\definecolor{mycolor4}{rgb}{0.87059,0.49020,0.00000}%
\definecolor{mycolor5}{rgb}{0.00000,0.44700,0.74100}%
\definecolor{mycolor6}{rgb}{0.74902,0.00000,0.74902}%

\newenvironment{shortlist}%
{\begin{list}{$\bullet$}
   {\setlength{\itemsep}{0in}
    \setlength{\labelsep}{0.05in}
    \setlength{\labelwidth}{0.2in}
    \setlength{\leftmargin}{0.25in}
    \setlength{\rightmargin}{0in}
    \setlength{\topsep}{0in}
   }}
{\end{list}}
\newcommand{\T}{^{\mbox{\tiny T}}}

\newcommand{\mcE}{\mathcal{E}}
\newcommand{\mcF}{\mathcal{F}}

\title{Approximate Support Recovery using Codes for Unsourced Multiple Access}
\author{Michail Gkagkos, Asit Kumar Pradhan, Vamsi Amalladinne, \\
Krishna Narayanan, Jean-Francois Chamberland, Costas N. Georghiades \\
Department of Electrical and Computer Engineering, Texas A\&M University}
\begin{document}

\maketitle

\begin{abstract}
We consider the approximate support recovery (ASR) task of inferring the support of a $K$-sparse vector $\xv \in \mathbb{R}^n$ from $m$ noisy measurements.
We examine the case where $n$ is large, which precludes the application of standard compressed sensing solvers, thereby necessitating solutions with lower complexity. 
We design a 
scheme for ASR by leveraging techniques developed for unsourced multiple access. 
We present two decoding algorithms with computational complexities 
$\mathcal{O}(K^2 \log n+K \log n \log \log n)$ and
$\mathcal{O}(K^3 +K^2 \log n+ K \log n \log \log n)$ per iteration, respectively.
When $K \ll n$, this is much lower than the complexity of approximate message passing with a minimum mean squared error denoiser
, which requires $\mathcal{O}(mn)$ operations per iteration.
This gain comes at a slight performance cost.
Our findings suggest that notions from multiple access 
can play an important role in the design of measurement schemes for ASR.
\end{abstract}




\section{Introduction and Problem Statement}

This article focuses on approximate support recovery,
a task that consists of estimating the locations of non-zero entries in an unknown sparse vector $\xv$ from measurements of the form
\begin{equation} \label{eqn:CS}
\yv = \mathbf{\Phi} \xv +  \wv 
\end{equation}
where $\mathbf{\Phi} \in \mathbb{R}^{m \times n}$ denotes a measurement (sampling) matrix and $\wv \in \mathbb{R}^m$ is additive noise.
The entries in $\wv$ consist of independent Gaussian random variables, each with mean zero and variance ${1}/{\mathsf{snr}}$, where $\mathsf{snr}$ is a fixed positive constant.
The unknown vector $\xv$ is $K$-sparse, i.e., $\|\xv\|_0 = K$; and the value of $K$ is known both to the encoder and decoder.
The objective is to infer the support of $\xv$ from observation $\yv$.
Variants of this problem have been studied extensively in the literature, e.g., ~\cite{candes2006stable,gilbert2007one,aeron2010information,zhang2011sparse,reeves2012sampling,tulino2013support,scarlett2016limits,gilbert2017all,chen2018sparse,li2019sub,inan2019group,calderbank2020chirrup}.
Herein, we assume that the choice of $\mathbf{\Phi}$ is entirely under the control of the designer.
We study the following two sparse models (SM) for $\xv$:
\begin{shortlist}
\item \textbf{SM1:} Non-zero entries in $\xv$ are equal to $\sqrt{{n}/{K}}$;
\item \textbf{SM2:} Non-zero entries in $\xv$ are independent Gaussian variables with mean zero and standard deviation 
$\sqrt{{n}/{K}}$.
\end{shortlist}
In our proposed scheme, we obtain measurement matrix $\mathbf{\Phi}$ by choosing one realization from a carefully crafted ensemble.
This selection process is independent of sparse vector $\xv$ and measurement noise $\wv$.
The rows of $\mathbf{\Phi}$ are normalized to each have vector norm one in expectation. 
For instance, when the entries of $\mathbf{\Phi}$ are i.i.d.\ random variables with mean zero, then their variance should be ${1}/{n}$.
 
We quantify the performance of a recovery algorithm using the empirical average miss detection $E_M$ and the average false alarm $E_F$;
these events can be expressed mathematically as
\begin{xalignat*}{2}
E_M &= \frac{1}{|\mathcal{K}|} \sum_{i=1}^n \mathbbm{1}_{\big\{i \in \mathcal{K},i \notin \hat{\mathcal{K}} \big\}} &
E_F &= \frac{1}{|\hat{\mathcal{K}}|} \sum_{i=1}^n \mathbbm{1}_{\big\{i \in \hat{\mathcal{K}},i \notin \mathcal{K} \big\}}
\end{xalignat*}
where $\mathcal{K}$ is the true support of $\xv$, $\mathcal{K} = \{i \in [n] : x_i \neq 0\}$, and $\hat{\mathcal{K}}$ denotes the estimated support procured by the recovery algorithm.
A related performance criterion we are interested in is based on the average mismatch 
$D= \max ( E_M, E_F )$.
 
In~\cite{reeves2012sampling}, Reeves and Gastpar derive information-theoretic bounds for the ASR problem in the linear sparsity regime where $K$ and $m$ grow linearly with $n$.
That is, $\lim_{n \rightarrow \infty} \frac{K}{n} = \kappa$ and $\lim_{n \rightarrow \infty}\frac{m}{n} = \rho$ for fixed constants $\kappa, \rho \in (0,1)$.
The quantities $\kappa$ and $\rho$ are called the sparsity rate and sampling rate, respectively.
Moreover, the authors also establish a lower bound on $\rho$ for given $\kappa$ and $D$.
They offer upper bounds (achievability results) on $\rho$ when $\mathbf{\Phi}$ is chosen with i.i.d.\ Gaussian entries.
Specifically, they list achievability results for three decoders: maximum likelihood (ML), approximate message passing (AMP) with soft thresholding or minimum-mean squared error (MMSE) denoiser, and linear MMSE (LMMSE) decoder.
The ML decoder has a complexity that is exponential in $K$ and, therefore, it is computationally impractical for most scenarios.
AMP decoders require at least $mn$ computations per iteration; and they substantially outperform the LMMSE decoder, whose complexity is $\mathcal{O}(n^3)$. 
For large dimensions, $\mathcal{O}(mn)$ remains a substantial computational challenge.
The design of schemes with lower computational complexity is therefore warranted.
This serves as a strong motivation for our work.

\subsection{Main Contributions} 

This article leverages recent advances in unsourced multiple access \cite{polyanskiy2017perspective,vem2019user,pradhan2020polar,amalladinne2020coded} to create an ensemble of measurement matrices that admits very low decoding complexity, sub-linear in $n$.
We present two versions of the decoder, which we label MF-SIC-MAP and MF-SIC-LS.
The computational complexity of the MF-SIC-MAP decoder is only $\mathcal{O}(K^2 \log n+K \log n \log \log n)$ per iteration; whereas the complexity of the MF-SIC-LS decoder is $\mathcal{O}(K^3 + K^2 \log n+K \log n \log \log n)$ per iteration. 
When $K = o(n^{\delta})$ with $\delta < 1/2$, the decoding complexity of the MF-SIC-LS is order-wise better than that of AMP. 
When $\delta \geq 1/2$, AMP has order-wise better complexity per iteration, yet the constants in our algorithm are substantially better than those of AMP for small $\kappa$, which is typically the regime of interest.
We show that, the sampling rate for the proposed scheme is only slightly worse than that of choosing $\mathbf{\Phi}$ from the i.i.d.\ Gaussian ensemble and decoding $\yv$ with the AMP algorithm.

Our proposed algorithmic architecture is inspired by a coding scheme that was recently published in \cite{pradhan2019polar} for a model akin to \textbf{SM1}.
There are several extensions beyond this work.
Firstly, we consider the matched filter (MF) receiver with serial interference cancellation instead of the MMSE receiver in \cite{pradhan2019polar}.
Secondly, we present a new asymptotic analysis of our ASR scheme for the \textbf{SM1} model, under the MF-SIC algorithm and for a single iteration; and we show that recovery can be performed with sub-linear complexity in $n$ for an appropriate choice of parameters. 
Using asymptotic analysis as a guide, successive interference
cancellation (SIC) and an estimator for the non-zero values are added to the overall scheme to improve performance for the \textbf{SM2} model.

\subsection{Unsourced Random Access \& Compressed Sensing}
The connection between multiple access and sparse recovery has been recognized for a long time \cite{wolf1985born}, \cite{inan2019group}. Most relevant to this paper is Unsourced Random Access (URA), which is a novel multi-user communication paradigm put forth by Polyanskiy~\cite{polyanskiy2017perspective} to meet the demands associated with massive connectivity in next-generation wireless networks.

\begin{figure}[h]
    \centering
    \scalebox{0.6}{\begin{tikzpicture}[
  font=\small, >=stealth', line width=0.75pt,
  block/.style={rectangle, draw, minimum height=7mm, minimum width=10mm, rounded corners},
symbol0/.style={rectangle, draw, fill=white, inner sep=0pt, minimum size=2.5mm},
symbol1/.style={rectangle, draw, fill=blue!50, inner sep=0pt, minimum size=2.5mm},
symbol2/.style={rectangle, draw, fill=white, inner sep=0pt, minimum size=2.5mm}
]

\
\node[block] (te1) at (-0.175,0) {$\mathcal{E}$};
 \node[block] (te2) at (-0.175,-1.25) {$\mathcal{E}$};
 \node (tei) at (-0.175,-2) {$\vdots$};
\node[block] (teKa) at (-0.175,-3) {$\mathcal{E}$};

\node[draw,circle,inner sep=3pt] (MAC) at (2.45,-1.5) {$\sum$};
\draw[->] (te1.east) -- (1.7,0) -- (MAC);
\draw[->] (te2.east) -- (1.7,-1.25) -- (MAC);
\draw[->] (teKa.east) -- (1.7,-3) -- (MAC);
\node (noise) at (2.45,-3.25) {Noise}
  edge[->] node[right]{$\ensuremath{\wv} \in \mathbb{R}^m$} (MAC);

\node[block,rotate=90,minimum width=25mm] (csDecoder) at (5.2,-1.5) {Decoder}
  edge[<-] (MAC);

\node(y) at (3.5,-0.75) {$\ensuremath{\yv} = \displaystyle \sum_{k=1}^{K} \ensuremath{\boldsymbol\phi_k x_k} + \ensuremath{\wv}$};
\node (v1) at (7.95,-1.5) {};
\draw[->]  (csDecoder) edge (v1);

\node at (1,-0.35) {$\boldsymbol\phi_3 \in \mathbb{R}^m$};
\node at (7.2,-1.15) {$\mathcal{L}(\ensuremath{\underline{y}}) = \{\hat{\ensuremath{\mv}}_1, \cdots, \hat{\ensuremath{\mv}}_K\}$};

\node[symbol0] (s05) at (-4,-4.00) {};
\node[symbol0] (s06) at (-4,-3.75) {};
\node[symbol0] (s07) at (-4,-3.50) {};
\node[symbol0] (s08) at (-4,-3.25) {};
\node[symbol1] (s09) at (-4,-3.00) {};
\node[symbol0] (s10) at (-4,-2.75) {};
\node[symbol0] (s11) at (-4,-2.50) {};
\node[symbol0] (s12) at (-4,-2.25) {};
\node[symbol0] (s13) at (-4,-2) {};
\node[symbol0] (s14) at (-4,-1.75) {};
\node[symbol0] (s15) at (-4,-1.50) {};
\node[symbol1] (s16) at (-4,-1.25) {};
\node[symbol0] (s17) at (-4,-1) {};
\node[symbol0] (s18) at (-4,-0.75) {};
\node[symbol0] (s19) at (-4,-0.50) {};
\node[symbol0] (s20) at (-4,-0.25) {};
\node[symbol1] (s21) at (-4,0.00) {};
\node[symbol0] (s22) at (-4,0.25) {};
\node[symbol0] (s23) at (-4,0.50) {};
\node[symbol0] (s24) at (-4,0.75) {};

\node at (-4,-4.5) {$\xv \in \mathbb{R}^n$};

\node[symbol0] (r1) at (-3.0,0.00) {};
\node at (-3.0,0.00){$0$};
\node[symbol0] (r2) at (-2.75,0.00) {};
\node at (-2.75,0.00){$0$};
\node[symbol0] (r3) at (-2.5,0.00) {};
\node at (-2.5,0.00){$0$};
\node[symbol2] (r4) at (-2.25,0.00) {};
\node at (-2.25,0.00){$1$};
\node[symbol2] (r5) at (-2.0,0.00) {};
\node at (-2.0,0.00){$1$};
\draw[->]  (s21) edge (r1);
\draw[->]  (r5) edge (te1);

\node[symbol0] (r6) at (-3.0,-1.25) {};
\node at (-3.0,-1.25){$0$};
\node[symbol2] (r7) at (-2.75,-1.25) {};
\node at (-2.75,-1.25){$1$};
\node[symbol0] (r8) at (-2.5,-1.25) {};
\node at (-2.5,-1.25){$0$};
\node[symbol0] (r9) at (-2.25,-1.25) {};
\node at (-2.25,-1.25){$0$};
\node[symbol0] (r10) at (-2.0,-1.25) {};
\node at (-2.0,-1.25){$0$};
\draw[->]  (s16) edge (r6);
\draw[->]  (r10) edge (te2);

\node[symbol0] (r11) at (-3.0,-3) {};
\node at (-3.0,-3){$0$};
\node[symbol2] (r12) at (-2.75,-3) {};
\node at (-2.75,-3){$1$};
\node[symbol2] (r13) at (-2.5,-3) {};
\node at (-2.5,-3){$1$};
\node[symbol2] (r14) at (-2.25,-3) {};
\node at (-2.25,-3){$1$};
\node[symbol2] (r15) at (-2.0,-3) {};
\node at (-2.0,-3){$1$};
\draw[->]  (s09) edge (r11);
\draw[->]  (r15) edge (teKa);

\node at (-2.5,0.35) {$\ensuremath{\mv}_1 \in \mathbb{F}_2^{\log n}$};
\node at (-2.5,-0.9) {$\ensuremath{\mv}_2$};
\node at (-2.5,-2.65) {$\ensuremath{\mv}_K$};
\node at (-4.5,0) {$3$};
\node at (-4.5,-1.25) {$8$};
\node at (-4.5,-3) {$15$};
\end{tikzpicture}}
    \caption{This figure illustrates the URA paradigm. Payload corresponding to active user $i$ is denoted by $\mv_i$ and its decimal representation is an element of the support of sparse vector $\xv$. All the active users encode their payloads into vectors of length $m$ using the same encoder $\mathcal{E}$.}
    \label{CS2MACschematic}
\end{figure}
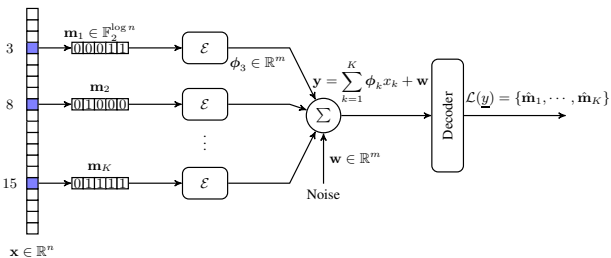

In the URA setting, $K$ active users in a network simultaneously transmit payloads of size $\log_2 n$ bits to an access point.
The destination is then tasked with recovering the list of payloads sent by these active users.
To this end, the message corresponding to every active user is encoded by an encoding function
$\mathcal{E}:\mathbb{F}_2^{\log_2 n} \rightarrow \mathbb{R}^m$
into a signal of length $m$, which is transmitted over a shared real-adder channel (see Fig.~\ref{CS2MACschematic}).
All the active users share a same codebook to encode their payloads.

In~\cite{polyanskiy2017perspective}, Polyanskiy points out that recovering the collection of sent messages (unordered) entails finding the support of a $K$-sparse vector of length $n$ from $m$ noisy measurements.
Thus, the signal received at the access point is analogous to \eqref{eqn:CS}, and the decoder is tasked with finding the support of $\xv$ given $\yv$, knowing $\mathbf{\Phi}$.
In a typical URA scenario, active users transmit payloads of size roughly $100$ bits, which implies that the length of $\xv$ is around $2^{100}$.
The sheer dimensionality of this support recovery problem precludes the usage of \textit{off-the-shelf} CS solvers.
Rather, it necessitates the design of novel, ultra-low complexity recovery algorithms.
The original URA formulation~\cite{polyanskiy2017perspective} is characterized by additive white Gaussian noise (AWGN) and signal amplitudes akin to model~\textbf{SM1} above.
Yet, this problem has also been extended to situations where active users experience fading~\cite{kowshik2019fundamental}.
In this latter case, the non-zero entries of $\xv$ correspond to different fading coefficients, a variation much close to model~\textbf{SM2} in spirit.
In addition to novel problem formulations, \cite{polyanskiy2017perspective, kowshik2019fundamental} offer achievability benchmarks for URA in the absence of complexity constraints.

Thenceforth, there has been significant research interest in the design of practical coding schemes that exhibit low decoding complexity and perform close to these achievability benchmarks. 
The coding schemes developed for URA can be broadly categorized into two groups: schemes built on traditional channel codes (e.g., \cite{ordentlich2017low,vem2019user,pradhan2019sparseidma,pradhan2020polar}),
and schemes that utilize the coded compressed sensing (CCS) framework (e.g., \cite{amalladinne2020coded,calderbank2020chirrup,Giuseppe,amalladinne2020unsourced}).

It is pertinent to note that a URA scheme can be transposed into a candidate solution for the ASR problem in \eqref{eqn:CS}.
The main idea is to consider $\mv_i$, the binary representation of integer~$i$, as the payload of an active user.
This is equivalent to constructing $\mathbf{\Phi}$ such that $\mathbf{\Phi}_{:,i} = \mathcal{E}(\mv_i)$ where $\mathbf{\Phi}_{:,i}$ is the $i$th column of $\mathbf{\Phi}$.
This construction shares some similarities to the construction of testing matrices 
for group testing in \cite{bondorf2020sublinear}. 
However, there are major differences too; our recovery scheme including the sequence detection stage, SIC stage
and the analysis are new and form substantially departure from published results.

\section{Design of the Sampling Matrix \& Encoding}

Having highlighted the connection between URA and the ASR problem, we now discuss the details of the encoding scheme $\mathcal{E}$.
Our scheme builds on and extends techniques from \cite{pradhan2019polar}, which utilizes
random spreading and single-user polar decoding as means to devise an ultra-low complexity URA algorithm with state-of-the-art performance.


\subsection{Multiple Access with Spreading Sequences}

A key concept in~\cite{pradhan2020polar} is to statistically separate the transmissions of $K$ active users with code division multiple access.
Following the URA designation in Fig.~\ref{CS2MACschematic}, we denote the binary representation of active index $k$ within sparse vector $\xv$ by $\mv_k$.
The length of compact vector $\mv_k$ is $B = \log_2 n$ bits.
This message is split into two parts: $\mv_{k,f}$ and $\mv_{k,s}$ of lengths $B_f$ and $B_s$, respectively, with $B = B_f + B_s$.
Based on the decimal representation of $\mv_{k,f}$, index~$k$ chooses one of the $J = \alpha K$ columns of the spreading dictionary $\mathbf{A}_t\in \{\pm \sqrt{1/n}\}^{L \times J}$, where $L =\beta K$, and $\beta$ is a constant. The entries of $\mathbf{A}_t $ are drawn independently from the set $\{\pm \sqrt{1/n}\}$ with equal probability.
The actual spreading operation for the $t^{th}$ coded bit is described in the next section. 
Additionally, the decimal representation of $\mv_{k,f}$ is employed to pick the positions of the frozen bits for polar encoding.

\subsection{Polar Encoder and Modulator}
To facilitate list decoding, the second part of $\mv_k$ is first padded with $r$ cyclic redundancy check (CRC) bits resulting in a message length of $r+B_s$ bits. 
Then a polar encoder maps this CRC augmented sequence into a codeword $\cv_{k}$ of length $n_c$.
Each coded bit of $c_{t,k}$, $t \in [1:n_c]$ is then BPSK modulated, and mapped to a symbol $b_{t,k} \in \{ -1,+1\}$.
Finally, every modulated bit $b_{t,k}$ acts as multiplicative factor for the spreading sequence $\av_{t,k}$, which was identified by the first part of the message $\mv_{k,f}$.
Given this encoding structure, we can see that the measurement matrix $\mathbf{\Phi}$ is composed of columns of the form
\begin{align*}
\begin{bmatrix}
b_1 \av_{1,j}\T & b_2 \av_{2,j}\T &
\cdots & b_{n_c} \av_{n_c,j}\T
\end{bmatrix}\T
\end{align*}
where $j = 1, 2, \ldots, J$ and $(b_1, b_2, \ldots, b_{n_c})$ is a valid BPSK modulated codeword of the corresponding polar code.
\begin{figure}[h]
    \centering
    \scalebox{0.65}{\begin{tikzpicture}
  [
  font=\normalsize, draw=black, line width=1pt,
  block/.style={rectangle, minimum height=10mm, minimum width=15mm,
  draw=black, fill=gray!10, rounded corners},
  ED/.style={rectangle, minimum height=10mm, minimum width=15mm, draw=black, rounded corners},
  message/.style={rectangle, minimum height=5.5mm, minimum width=15mm, draw=black, rounded corners}
  ]


\node[block,align=center] (idx)  {Index};
\node[block,right of= idx, node distance = 2.5cm] (dectobin) {DecToBin};
\draw[->] (idx) -- (dectobin) node[midway,sloped,above] {$k$};
\node[block,right of= dectobin, node distance = 2.5cm] (msg)  {Message};
\draw[->] (dectobin) -- (msg) node[midway,sloped,above]{${\bf m}_k$};

\node[block,below of= msg, node distance = 1.8cm] (polar){Polar Encoder};
\draw[->] (msg) -- (polar) node[midway,sloped,right,rotate=90]{${\bf m}_{k,s}$};
\node[block,right of= msg, node distance = 3.0cm] (bintodec) {BinToDec};
\draw[->] (msg) -- (bintodec) node[midway,sloped,above]{${\bf m}_{k,f}$};
\node[block,below of= polar, node distance = 1.8cm] (bpsk){BPSK};
\draw[->] (polar) -- (bpsk)node[midway,sloped,right,rotate=90]{$ c_{1:nc,k}$};
\draw[->] (bintodec) -- (polar) node[midway,sloped,above]{$j$};

\node[block,below of= bintodec, node distance = 1.8cm] (A){$A$};
\draw[->] (bintodec) -- (A)node[midway,sloped,right,rotate=90]{$j$};

\node[block,right of= A, node distance = 3.0cm] (value){Value};

\node[block,below of= A, node distance = 1.8cm] (phi){$\boldsymbol\phi_k$};
\node[block,below of= value, node distance = 1.8cm] (output){$\boldsymbol\phi_kx_k$};
\draw[->] (A) -- (phi)node[midway,sloped,right,rotate=90]{$\{{\bf a}_{t,j}\}_{t=1}^{n_c}$};
\draw[->] (bpsk) -- (phi)node[midway,sloped,above]{$b_{1:nc,k}$};
\draw[->] (phi) -- (output);
\draw[->] (value) -- (output)node[midway,sloped,right,rotate=90]{$x_k$};


  

\end{tikzpicture}}
    \caption{This block diagram offers a synopsis of the encoding process or, equivalently, it is an implicit construction for the sampling matrix.}
    \label{fig:encoder}
\end{figure}
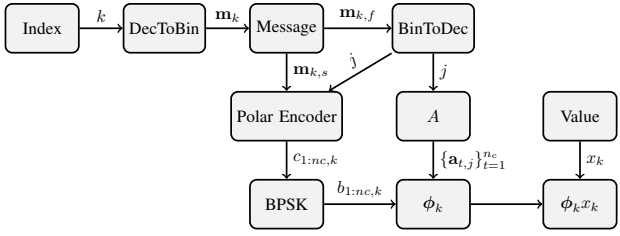

\section{Recovery Algorithm}
The proposed iterative recovery algorithm consists of three main components; a  sequence detector, which identifies the spreading sequences picked by active indices; a polar decoder; and a successive interference canceller.
The sequence detection is based on a combination of matched filtering (MF) and hypothesis testing (HT).
If the unknown vector $\xv$ is drawn from model \textbf{SM2}, then there is an additional operation to estimate the amplitudes of non-zero values in $\xv$.
The SIC removes the contribution of successfully decoded signals, and feeds the residual back to the sequence detector.


\subsection{Spreading Sequences Detector}

The received signal $\yv$ is reshaped in a form amenable to sequence detection as follows.
\begin{align}
    \mathbf{\bar{Y}} =
    \begin{bmatrix}
    \bar{\yv}_1 & \bar{\yv}_2 & \dots & \bar{\yv}_{n_c}
    \end{bmatrix}
\end{align}
where $\bar{\yv}_t \in \mathbb{R}^L$ and $L = \beta K$.
Section $\bar{\yv}_t$ denotes the received signal corresponding to the $t^{th}$ polar coded symbol.
Since the spreading sequences are picked by active indices based on a portion of their own binary representations, multiple indices can choose a same spreading sequence.
Parameter $B_f$ is chosen sufficiently large to ensure that such collisions have a low probability.
For ease of exposition, we assume there are no collisions in the remainder of the description.
Yet, the scheme does not rely on this assumption; operations can be modified to accommodate the more general case where multiple indices can choose a same spreading sequence.
Throughout the discussion, we assume the MF is acting on the latest sketch afforded by the SIC update.
This latest sketch consists of the residual in $\mathbf{\bar{Y}}$, with the contributions of the decoded non-zero entries subtracted from the original measurement vector.
Assuming perfect interference cancellation, the $t^{th}$ column of the updated  sketch $\mathbf{\bar{Y}}^{(s)}$  can be expressed as
\begin{equation}
\textstyle
\bar{\yv}_t^{(s)} =
\sum_{k \in \mathcal{K} \setminus \mathcal{K}^{(s)}} \av_{t,k} b_{t,k}x_k  
 + \bar{\wv}_t
\end{equation}
where $\mathcal{K}^{(s)}$ denotes the collection of indices that correspond to the $s$ subtracted codewords.
The sequence detector correlates $\bar{\yv}_t^{(s)}$ with every column of $\mathbf{A}_t^{(s)}$,
\begin{equation} 
\label{eq:BefZ}
\begin{split}
\big\langle \av_{t,k}, \bar{\yv}_t^{(s)} \big\rangle
&= (L/n)x_k b_{t,k} + \langle \av_{t,k}, \bar{\wv}_t \rangle \\
&\quad + 
\textstyle \sum_{k' \in \mathcal{K}^{(s)} \setminus k} {b_{t,k'}x_{k'}} \langle \av_{t,k}, \av_{t,k'} \rangle .
\end{split}
\end{equation}
The scaled output of these operations serve as a test statistic to determine whether sequence $\av_{:,k}$ is active.
The elements of this test statistic take the form
\begin{equation}
\begin{split}
Z_{t,k} &= \big\langle \av_{t,k}, \bar{\yv}_t^{(s)} \big\rangle \sqrt{n} / \sqrt{L} \\
&= \sqrt{\beta \kappa}x_k b_{t,k}  + I_{t,k} + N_{t,k} .
\end{split}
\end{equation}
We note that $Z_{t,k}$, $I_{t,k}$, and $N_{t,k}$ implicitly depend on $s$; but their superscripts are omitted for notational convenience.
Through the central limit theorem, we can approximate the effect of interference $I_{t,k}$ by a zero-mean Gaussian random variable with variance $\frac{K-1-s}{K}$.
As a result, \eqref{eq:BefZ} can be viewed as a shifted-mean Gaussian HT problem with effective noise $V_{t,k} = I_{t,k} + N_{t,k}$.
The goal of the spreading sequence detector is to use  $Z_{1:n_c,k} \triangleq ( Z_{1,k},\dots,Z_{n_c,k} )$ to decide whether $\av_{:,k}$ is active or not.
Let $S_k$ denote the event that $\av_{:,k}$ is active. 
The prior probability of this event can be computed as $\Pr(S_k) = 1 - (1 - {1}/{J})^{K-s}$.
We note, briefly, that the form of the MAP decision rule depends on the signal model adopted.
Below, we describe the decision rule for non-zero signals drawn from \textbf{SM2}.
The HT problem for \textbf{SM1} can be formulated in a similar manner; it admits a simpler form and, as such, details are omitted.
Under \textbf{SM2}, when $\av_{:,k}$ is active, the output of the $t^{th}$ MF is given by
\begin{equation}
Z_{t,k} = \sqrt{\beta \kappa} x_k b_{t,k} + V_{t,k}
\end{equation}
where $V_{t,k} \sim \mathcal{N} \left( 0,\frac{K-1-s}{K} + \frac{1}{\mathsf{snr}} \right)$.
On the other hand, if $\av_{:,k}$ is inactive, then $Z_{t,k} = {V}_{t,k}$, where ${V}_{t,k} \sim \mathcal{N} \left( 0,\frac{K-s}{K}+\frac{1}{\mathsf{snr}} \right)$.
Altogether, when $\xv$ is drawn from \textbf{SM2}, $Z_{t,k}$ is distributed as 
\begin{equation*}
Z_{t,k} \sim
\begin{cases}
\mathcal{N} \left( 0, \beta + \frac{K-1-s}{K} + \frac{1}{\mathsf{snr}} \right) & \av_{:,k} \text{ is active} \\
\mathcal{N} \left( 0,\frac{K-s}{K} + \frac{1}{\mathsf{snr}} \right) & \text{otherwise} .
\end{cases}
\end{equation*}
The log-likelihood ratio (LLR) corresponding to the activity of sequence $\av_{:,k}$ is then given by
\begin{equation} \label{eq:LLR-ED}
\log \frac{ \Pr( S_k | Z_{1:n_c,k})}
{\Pr( \bar{S}_k | Z_{1:n_c,k})}
= \log\frac{\Pr( S_k ) f( Z_{1:n_c,k} | S_k)}
{\Pr( \bar{S}_k ) \prod_{t=1}^{n_c} f( Z_{t,k} | \bar{S}_k)}
\end{equation}
where $f(Z_{1:n_c,k}|S_k)$ is the joint pdf of sequence $Z_{1:n_c,k}$ given  $S_k$, and $f(Z_{t,k}|\bar{S}_k)$ is the pdf of a Gaussian distribution with mean 0 and variance $\frac{K-s}{K}+\frac{1}{\mathsf{snr}}$.
The quantity $f(Z_{1:n_c,k}|S_k)$ in \eqref{eq:LLR-ED} can be evaluated numerically.
Ultimately, $\av_{:,k}$ is deemed active if the LLR in \eqref{eq:LLR-ED} is greater than a threshold $\gamma$.

Again, a similar HT problem can be formulated for \textbf{SM1}.
Interestingly, in this alternate case, the conditional distributions of the LLR are tractable.
In Sec.~\ref{sec:Analysis}, we formulate an optimization framework to choose system parameters based on the analysis afforded by this more accessible setting.

\subsection{Detection of Polar Codewords}

In this section, we describe the detection of valid polar codewords.
The key idea stems from the realization that, if spreading sequence $\av_{:,k}$ is active, then the elements of $Z_{1:n_c,k}$ can act as estimates of the polar coded bits $b_{1:n_c,k}$.
Yet, when the sparse vector $\xv$ is drawn from \textbf{SM2} model, the sign of $x_k$ is unknown and there is a need to run two list decoders.
The inputs to these two list decoders are $Z_{1:n_c,k}$ and $-Z_{1:n_c,k}$, respectively.
On the other hand, if $\xv$ is drawn from \textbf{SM1}, we only need one decoder with an input $Z_{1:n_c,k}$.
The list decoder verifies CRC constraints for every decoded codeword.
If two or more codewords satisfy the checks, the most likely codeword is passed to the next step.
Finally, a hard decision decoder is applied to $Z_{1:n_c,k}$ and its output is compared to the output of the list decoder(s).
If the two bit streams differ in more than a few positions, the codeword is discarded.

\subsection{Estimation of Non-Zero Entries in $\xv$ }

During every SIC iteration, once non-zero locations are identified, their values may need to be estimated.
This is unnecessary for \textbf{SM1} because non-zero amplitudes are known and equal to $\sqrt{{n}/{K}}$.
But this step is crucial when $\xv$ is drawn from \textbf{SM2} in order to facilitate SIC.
This estimation process can be accomplished using standard techniques and, in general, estimators with higher computational cost offer better performance.
In our numerical results, we investigate the performance of two estimators; a maximum a posteriori (MAP) estimator and a least squares (LS) estimator.
The latter is more complex, yet it exhibits a better performance.

\subsection{Successive Interference Cancellation}

The contributions of the recovered non-zero entries in the sparse signal are removed from the received signal in the spirit of SIC.
The residual is then passed to the sequence detector for the next decoding round.
This process continues until all the transmitted
messages are recovered successfully or there is no improvement between two consecutive rounds of iterations. 
The estimated support output by the algorithm corresponds to the locations of $K$ largest entries in absolute value of the estimated signal.
The rules of SIC are fairly standard and we omit the details due to space constraints.

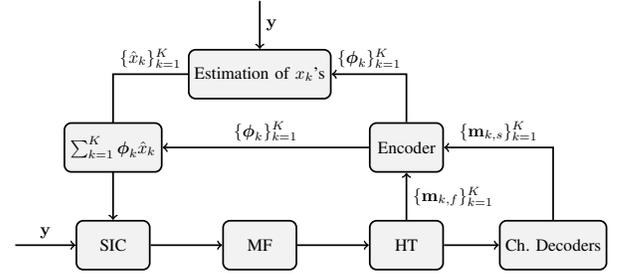
\begin{figure}[h]
    \centering
    \scalebox{0.65}{\begin{tikzpicture}
  [
  font=\normalsize, draw=black, line width=1pt,
  block/.style={rectangle, minimum height=10mm, minimum width=15mm,
  draw=black, fill=gray!10, rounded corners},
  sum/.style={rectangle, minimum height=15mm, minimum width=15mm, ddraw=black, rounded corners, fill=gray!10},
  message/.style={rectangle, minimum height=5.5mm, minimum width=15mm, draw=black, rounded corners}
  ]

\node[block,align=center] (sic)  {SIC};
\node[block,right of= sic, node distance = 3cm] (mf) {MF};

\draw[->] (sic) -- (mf);

\node[block,right of= mf, node distance = 3cm] (LLR) {HT};

\draw[->] (mf) -- (LLR);


\node[block,right of= LLR, node distance = 3cm] (cd) {Ch. Decoders};
\draw[->] (LLR) -- (cd);


\node[block,above of= LLR, node distance = 2cm] (enc) {Encoder};
\draw[->] (LLR) -- (enc) node[midway,sloped,right,rotate=270] {$\{{\bf m}_{k,f}\}_{k=1}^K$};
\draw[->] (cd) |- (enc) node[near end,sloped,above] {$\{{\bf m}_{k,s}\}_{k=1}^K$};

\node[block,above of= mf, node distance = 3.5cm] (ls) { Estimation of $x_k$'s};
\draw[->](-2,0) -- (sic.west)  node[midway,above]{${\bf y}$};
\draw[->](enc) |- (ls)  node[near end,above]{$\{\boldsymbol\phi_k\}_{k=1}^K$};

\draw[->] (ls) -| (sic) node[near start,sloped,above] {$\{\hat{x}_k\}_{k=1}^K$};
\node[block,above of= sic, node distance = 2cm] (sum) {$\sum_{k=1}^K\boldsymbol\phi_k\hat{x}_k$};
\draw[->] (enc) -- (sum) node[midway,sloped,above] {$\{\boldsymbol\phi_k\}_{k=1}^K$};

\draw[->] (3,5) -- (ls.north)   node[midway,right]{${\bf y}$};
\end{tikzpicture}}
    \caption{This block diagram shows the recovery scheme.}
    \label{fig:decoder}
\end{figure}

\section{Analysis of Probability of Error for MF-SIC-Bernoulli}
\label{sec:Analysis}
In this section, we analyse the probability of error for the MF-SIC-Bernoulli approach without SIC and, based on this analysis, we pick parameters for our simulations.
When $\av_{:,k}$ is active, the sum of the squares of $Z_{1:n_c,k}$ possesses a (scaled) non-central Chi-square distribution.
Whereas when $\av_{:,k}$ is inactive, the sum of squares features a (scaled) central Chi-square distribution.
The parameters of these distributions are straightforward to obtain, but left out due to space limitation.
Let $T_k$ be the sum of squares described above.
This quantity is employed as a statistic to assess whether sequence $\av_{:,k}$ is active.
Specifically, we fix a threshold $\gamma$ and our sequence detector classifies $\av_{:,k}$ as active whenever
\[
T_k > (1+\gamma) n_c .
\]
Since the signal construction process is symmetric with respect to indices in $\xv$,
we can analyze the per-index probability of error by focusing exclusively on index~1.
Without loss of generality, suppose that index~1 maps to sequence $\av_{:,1}$.
Then, we can introduce the following error events.
\begin{shortlist}
\item[$\mathcal{E}_1$:] Active spreading sequence $\av_{:,1}$ is selected by another non-zero entry in $\xv$, on top of index~1.
\item[$\mathcal{E}_2$:] The sequence detector misclassifies $\av_{:,1}$ as inactive.
\item[$\mathcal{E}_3$:] The polar codeword produced by non-zero index~1 fails to be decoded.
\item[$\mathcal{F}$:] The recovery algorithm erroneously outputs index 1 when its true value is zero.
\end{shortlist}
Let us define the missdetection rate $P_M$ and probability of false alarm $P_F$ of the $i$ non-zero value as follows,
\begin{xalignat*}{2}
    P_M &= \Pr \big( i \notin \hat{\mathcal{K}}  |i \in \mathcal{K} \big) &
    P_F &= \Pr \big( i \in \hat{\mathcal{K}}|i \notin \mathcal{K} \big) .
\end{xalignat*}
Based on the events above, they can be upper bounded by
\begin{xalignat*}{2}
    P_M &\leq \Pr(\mcE_1) + \Pr(\mcE_2) + \Pr(\mcE_3) &
    P_F &\leq K(\alpha-1) \Pr(\mcF) .
\end{xalignat*}
$P_F$ comes from assuming that the decoder will eventually always output exactly $K$ indices as active.
To meet target error probability $\varepsilon$, our approach is to make sure that $\Pr(\mcE_i) \leq {\varepsilon}/{3}$ and $\Pr(\mcF) \leq \varepsilon$.

\subsection{Analysis of $\Pr(\mcE_1)$}

As part of the generation process, every index in $\xv$ gets mapped independently and uniformly to a spreading sequence.
This is an instance of the classic balls-and-bins problem.
There are $J = \alpha K$ spreading sequence and $K$ indices.
It follows that the probability of a bin collision for active index~1 is asymptotically given by
\begin{equation}
    \lim_{K \rightarrow \infty} \Pr(\mcE_1) = 1 -  e^{-\frac{1}{\alpha}} - {e^{-\frac{1}{\alpha}}}/{\alpha} .
\end{equation}
We can pick $\alpha$ such that $\Pr(\mcE_1) \leq {\varepsilon}/{3}$.

\subsection{Analysis of $\Pr(\mcE_2)$}
Let $F_{\chi^2}(x;\lambda,n_c)$ denote the cumulative distribution function of a $\chi^2$ random variable with non-centrality parameter $\lambda$ and $n_c$ degrees of freedom evaluated at $x$.
We pick a threshold for the sequence detector $\gamma$ such that
\begin{equation}
    \Pr(\mcE_2) = F_{\chi^2}(n_c(1+\gamma);\lambda,n_c) \leq \epsilon/3 . \label{eq:cntr1}
\end{equation}

\subsection{Analysis of $\Pr(\mcE_3)$}
We assume that the code used for each index achieves the finite block length bound for that length, rate, and equivalent SNR.
Let $P_{\sf{FBL}}(k_c,n_c,\sf{SNR_{eq}})$ denote the achievable probability of error for a code with dimension $k_c$, codeword length $n_c$, when used with an additive white Gaussian noise channel with signal to noise ratio $\sf{SNR_{eq}}$.
We select parameters such that
\begin{equation}
    \Pr(\mcE_3) \leq P_{\sf{FBL}}\bigg(\log \frac{n}{\alpha K} + l_{\sf{CRC}},n_c,\frac{\beta}{\frac{K-1}{K}+\frac{1}{\mathsf{snr}}}\bigg)\leq \frac{\epsilon}{3} . \label{eq:cntr2}
\end{equation}
While there is no closed-form expression for $P_{\sf{FBL}}$, this bound can be evaluated numerically.

\subsection{Analysis of $\Pr(\mcF)$}
A false alarm occurs when the sum of squares of an inactive sequence exceeds threshold $(1+\gamma)n_c$ and the candidate polar codeword fulfills the CRC.
Thus, we want
\begin{equation}
    \Pr(\mcF) = \big(1-F_{\chi^2}(n_c(1+\gamma);0,n_c)\big) \frac{1}{2^{l_{\sf CRC}}} \leq  \frac{\epsilon}{K(\alpha-1)}\label{eq:cntr3}
\end{equation}
The design problem can then be posed as choose parameters $\beta$, $n_c$, $l_{\sf CRC}$ to minimize $\beta n_c$, subject to constraints \eqref{eq:cntr1}, \eqref{eq:cntr2}, and \eqref{eq:cntr3}.
The problem can be solved numerically.

\section{Decoding Complexity}
In this section, we briefly describe the decoding complexity of the proposed scheme, which features a decoding algorithm that occurs in three steps.
In the first step, matched filters are employed to obtain estimates of the coded symbols. 
Since there are $n_c$ coded symbols and the length of each spreading sequence is $L$, its complexity is $\mathcal{O}(n_cL)$.
This step is followed by the list decoding of the polar codes whose complexity is $\mathcal{O}(n_c\log n_c)$. 
These  two  steps  need  to  be  repeated for each of the $J$ spreading sequences and for each of the non-zero entries in $\xv$, respectively; hence, the overall complexity of these two steps is $\mathcal{O}(Jn_cL+Kn_c \log n_c)$. 
In the last step, an LS estimator is employed to obtain the non-zero entries and this step has a complexity of $\mathcal{O}(K^3)$.
Since $L$ and $J$ are $\mathcal{O}(K),$ and $n_c$ is $\mathcal{O}(\log n),$ the overall decoding complexity is $\mathcal{O}(K^3+K^2 \log n+K \log n \log \log n)$.
Admittedly, this can be a concern when $K$ is large, but it is adequate for a range of practical sizes.
On the other hand, a MAP estimator instead of LS  results in a much lower decoding complexity that scales as $\mathcal{O}(K^2 \log n+K \log n \log \log n)$.
However, the computational complexity of AMP-MMSE scales as $\mathcal{O}(mn).$
When applied to very high dimensional signals and in the \textit{very sparse} regime, the complexity of the proposed scheme is substantially lower than that of AMP-MMSE.
\section{Numerical Results}
In this section, we study the performance of the proposed framework and compare it with a recovery algorithm that employs AMP with a separable MMSE denoiser.
The latter, which we term AMP-MMSE, has a much higher computational and storage complexity when compared to the proposed scheme.
The sparsity index of the unknown sparse vector $\xv$ is $K=25$ and the length is $n=2^{18}$.
The target error probability is set to $D = 0.07$ for all simulations.
Figure~\ref{fg:bernoulli} showcases the performance of the proposed scheme with and without SIC and juxtaposes it with that of AMP-MMSE when sparse vectors are drawn from the \textbf{SM1} model.
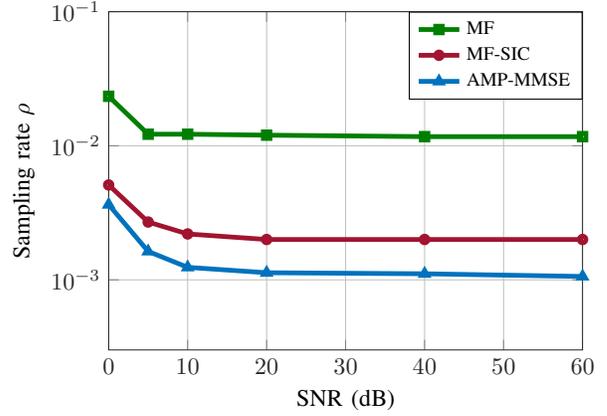
\begin{figure}[t!]
\centerline{\begin{tikzpicture}[scale=0.9]
\definecolor{mycolor1}{rgb}{0.63529,0.07843,0.18431}%
\definecolor{mycolor2}{rgb}{0.00000,0.44706,0.74118}%
\definecolor{mycolor3}{rgb}{0.00000,0.49804,0.00000}%
\definecolor{mycolor4}{rgb}{0.87059,0.49020,0.00000}%
\definecolor{mycolor5}{rgb}{0.00000,0.44700,0.74100}%
\definecolor{mycolor6}{rgb}{0.74902,0.00000,0.74902}%

\begin{axis}[%
font=\normalsize,
width=7cm,
height=5cm,
ymode=log,
scale only axis,
every outer x axis line/.append style={white!15!black},
every x tick label/.append style={font=\color{white!15!black}},
xmin=0,
xmax=60,
xtick = {0,10,20,30,40,50,60},
xlabel={SNR (dB)},
xmajorgrids,
xminorgrids,
every outer y axis line/.append style={white!15!black},
every y tick label/.append style={font=\color{white!15!black}},
ymin=0.0003,
ymax=0.1,
ytick = {0.1,0.01,0.001,...,0.001},,
ylabel={Sampling rate $\rho$},
ymajorgrids,
yminorgrids,
legend style={at={(1, 1)},anchor=north east,draw=black, fill=white, legend cell align=left,font=\footnotesize}]

\addplot [color=mycolor3,solid,line width=2.0pt,mark size=1.4pt,mark=square,mark options={solid}]
  table[row sep=crcr]{
0	0.0234
\\
5	0.0122
\\
10	0.0122
\\
20	0.0120
\\
40 0.0117
\\
60 0.0117\\
};
\addlegendentry{ MF};
\addplot [color=mycolor1,solid,line width=2.0pt,mark size=1.4pt,mark=o,mark options={solid}]
  table[row sep=crcr]{
0	0.0051\\
5	0.0027\\
10	0.0022\\
20	0.002\\
40	0.002\\
60	0.0020\\
};
\addlegendentry{MF-SIC};


\addplot [color=mycolor2,solid,line width=2.0pt,mark size=1.4pt,mark=triangle,mark options={solid}]
  table[row sep=crcr]{
0	3.64e-3
\\
5	1.63e-3
\\
10	1.24e-3
\\
20	1.13e-3
\\
40	1.11e-3
\\
60 1.06e-3 
\\
};
\addlegendentry{AMP-MMSE};

 \end{axis}

\end{tikzpicture}%
}
 \caption{This plot shows a performance comparison between the proposed schemes and AMP-MMSE when the sparse signals are generated under \textbf{SM1}.}
\label{fg:bernoulli}
\end{figure}
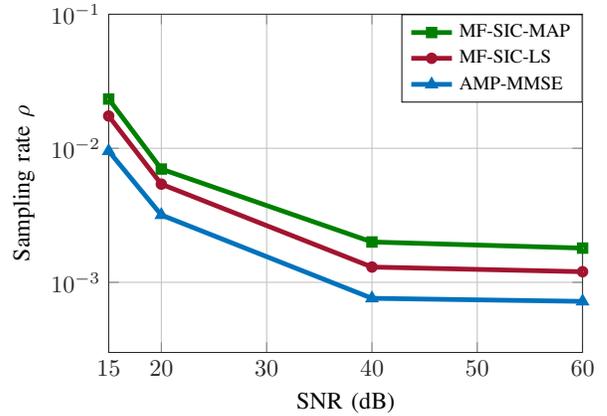
\begin{figure}[t!]
\centerline{\begin{tikzpicture}[scale=0.9]
\definecolor{mycolor1}{rgb}{0.63529,0.07843,0.18431}%
\definecolor{mycolor2}{rgb}{0.00000,0.44706,0.74118}%
\definecolor{mycolor3}{rgb}{0.00000,0.49804,0.00000}%
\definecolor{mycolor4}{rgb}{0.87059,0.49020,0.00000}%
\definecolor{mycolor5}{rgb}{0.00000,0.44700,0.74100}%
\definecolor{mycolor6}{rgb}{0.74902,0.00000,0.74902}%

\begin{axis}[%
font=\normalsize,
width=7cm,
height=5cm,
ymode=log,
scale only axis,
every outer x axis line/.append style={white!15!black},
every x tick label/.append style={font=\color{white!15!black}},
xmin=15,
xmax=60,
xtick = {15,20,30,40,50,60},
xlabel={SNR (dB)},
xmajorgrids,
xminorgrids,
every outer y axis line/.append style={white!15!black},
every y tick label/.append style={font=\color{white!15!black}},
ymin=0.0003,
ymax=0.1,
ytick = {0.1,0.01,0.001,...,0.0001},,
ylabel={Sampling rate $\rho$},
ymajorgrids,
yminorgrids,
legend style={at={(1, 1)},anchor=north east,draw=black, fill=white, legend cell align=left,font=\footnotesize}
]

\addplot [color=mycolor3,solid,line width=2.0pt,mark size=1.4pt,mark=square,mark options={solid}]
  table[row sep=crcr]{
5	0\\
15 0.0233\\
20	0.007\\
40	0.0020\\
60	0.0018\\
};
\addlegendentry{MF-SIC-MAP};
\addplot [color=mycolor1,solid,line width=2.0pt,mark size=1.4pt,mark=o,mark options={solid}]
  table[row sep=crcr]{
5	0\\
15 0.0174\\
20	0.0054\\
40	0.0013\\
60	0.0012\\
};
\addlegendentry{MF-SIC-LS};
\addplot [color=mycolor2,solid,line width=2.0pt,mark size=1.4pt,mark=triangle,mark options={solid}]
  table[row sep=crcr]{
15	0.00953
\\
20	3.18e-3
\\
40 7.6e-4
\\
60 7.21e-4
\\
};
\addlegendentry{AMP-MMSE};




\end{axis}

\end{tikzpicture}%
}
 \caption{This plot compares the performance of the proposed schemes and that of AMP-MMSE when the sparse signals are generated under the more intricate \textbf{SM2} model.}
\label{fg:gaussian}
\end{figure}

Figure~\ref{fg:gaussian} offers similar results when the sparse signals are drawn under \textbf{SM2}.
For this scenario, the performance of the proposed scheme with LS estimation, termed MF-SIC-LS, is very close to that of AMP-MMSE.
Employing a (marginal) MAP estimator instead of LS results in a much lower decoding complexity.
As anticipated, this gain comes with a slight loss in performance in terms of sampling rate.

\clearpage
\bibliographystyle{IEEEbib}
\bibliography{MACcollison}

\end{document}